\documentclass[aps,prl,twocolumn,floatfix]{revtex4}
\usepackage{amsmath}
\usepackage{amssymb}
\usepackage[dvips]{graphicx}

\newcommand{\up}{\uparrow}
\newcommand{\down}{\downarrow}
\newcommand{\dx}{\partial_x}

\newcommand{\arXiv}[1]{\texttt{arXiv:}\,#1}

\begin{document}

\title{Multi-particle composites in density-imbalanced quantum fluids}

\author{Evgeni Burovski}
\author{Giuliano Orso}
\author{Thierry Jolicoeur}
\affiliation{Laboratoire de Physique Th\'eorique et Mod\`eles statistiques,
Universit\'e Paris-Sud, 91405 Orsay, France}


\begin{abstract}
We consider two-component one-dimensional quantum gases with density
imbalance. While generically such fluids are two-component Luttinger liquids,
we show that if the ratio of the densities is a rational number, $p/q$,
\emph{and} mass asymmetry between components is sufficiently strong, one of the two
eigenmodes acquires a gap. The gapped phase corresponds to (algebraic)
ordering of $(p+q)$-particle composites. In particular, for attractive
mixtures, this implies that the superconducting correlations are destroyed. We
illustrate our predictions by numerical simulations of the fermionic Hubbard
model with  hopping asymmetry.
\end{abstract}


\maketitle

Thanks to recent advances in experimental techniques of dealing with cold
gases, it is now feasible to engineer one-dimensional (1D) quantum fluids
by confining atoms in cigar-shaped traps with tight radial confinement
\cite{expt1D}.
By devising an appropriate optical lattice it is also possible to construct a
weakly coupled array of such 1D ``tubes,'' thus allowing one to study the
dimensional crossover from 1D to three dimensions. A number of ongoing and planned
experiments deals with two-component mixtures atoms of either statistics, i.e.
Fermi-Fermi (FF), Bose-Bose (BB) or Bose-Fermi (BF) mixtures
\cite{PitaevskiiRMP2008}.
Most of recent theoretical work dealt with equal-density mixtures, where a
rich phase diagram containing both gapped and gapless phases was found
\cite{CazalillaHo2003,GiamarchiCazalillaHo2005,Mathey2007}. For mixtures with
unequal densities the ground state is generally found to be a two-component
Luttinger liquid \cite{CazalillaHo2003,MatheyWang2007,LuWangGuLin2009}. For
attractive FF mixtures, superconducting correlations dominate, thus making the
ground state a 1D analog of the long-elusive Fulde-Ferrell-Larkin-Ovchinnikov
(FFLO) phase, as confirmed both by  Bethe Ansatz calculations for integrable
models \cite{Orso2007} and numerical simulations
\cite{Feiguin2007,DasSarma2009}. The case of unequal mass mixtures---where
integrable microscopic models are not available---has been studied
analytically by means of effective field theory
\cite{CazalillaHo2003,LuWangGuLin2009}, and numerically by Monte Carlo
\cite{Batrouni2008} and time-evolving block decimation (TEBD) \cite{DasSarma2009} methods. A common result
which emerges is that for strong enough mass asymmetry and/or strong enough
attraction the system collapses, while for moderate mass asymmetry and
non-zero density imbalance the ground state is again a gapless two-component
Luttinger liquid with an FFLO-type algebraic order.

In this Letter we study a generic two-component 1D mixture with density
imbalance within the harmonic fluid approach (``bosonization''). We reveal a
generic mechanism which, for a certain relation between the densities, opens a
gap in the excitation spectrum and completely destroys superconducting
correlations. We concentrate on the properties of the FF mixtures, but our
predictions are applicable to BF and BB mixtures with minor modifications. Our
findings might also be relevant to spin ladder materials
with non-equivalent chains in high magnetic fields. We further corroborate our
predictions by DMRG simulations \cite{DMRG} of a Hubbard model with hopping
asymmetry.

Consider the mixture of two sorts of fermionic atoms, which
we label by  a pseudo-spin index $\sigma = \up, \down$. In the
bosonization approach we introduce for each species a pair of scalar fields
$\phi_\sigma(x)$ and  $\theta_\sigma(x)$ which vary slowly on the scale of
$n_\sigma^{-1}$, where $n_\sigma$ are the average densities
\cite{GiamarchiTheBook}.
Using the Haldane construction we write for the field operators
$\Psi^\dagger_\sigma(x) \sim \left( n_{\sigma}- \dx\phi_\sigma/\pi
\right)^{1/2} \sum_{s} e^{ i s (\pi n_\sigma x -\phi_\sigma ) } e^{-i
\theta_\sigma}$ where the summation over $s$ runs over odd integers $s$.
\cite{Haldane1981} For the density operator, $\hat{n}_\sigma$, this leads to~:
\begin{equation}
\hat{n}_\sigma(x) \sim \left( n_\sigma - \dx\phi_\sigma / \pi \right)%
\sum_s e^{ 2 i s (\pi n_\sigma x -\phi_\sigma) } \; . %
\label{HaldaneN} \\
%
\end{equation}
One of the advantages of the Haldane representation \eqref{HaldaneN} is that
an effective low-energy Hamiltonian can be written solely in terms of
$\phi_\sigma$ and $\Pi_\sigma$ \cite{GiamarchiTheBook}. In the non-interacting
case it is given by $\mathcal{H}_\mathrm{free} = \mathcal{H}_0(\phi_\up)
+\mathcal{H}_0(\phi_\down)$, where~:
\begin{equation}
%
\mathcal{H}_0(\phi_\sigma) = \dfrac{v_\sigma}{2\pi} \int\! dx\,\left[
K_\sigma(\pi\Pi_\sigma)^2 +K_\sigma^{-1} \left(\dx\phi_\sigma\right)^2 \right]
\; ,
\label{H0}
\end{equation}
where $v_\sigma$ are Fermi velocities and $K_\sigma = 1$ the so-called
Luttinger parameters equal to one in the free case. In presence of
density-density interactions, $\int\! dx dx'\, U_{\sigma\sigma'}(x-x')
\hat{n}_\sigma(x) \hat{n}_\sigma'(x')$, Eq.\ \eqref{H0} is modified in several
ways. First of all, the $s=0$ terms of Eq.\ \eqref{HaldaneN} give rise to an
acoustic coupling~:
\begin{equation}
\mathcal{H}_1 = g \int\! dx \, (\dx \phi_\up)(\dx \phi_\down) \; , \label{H1}
\end{equation}
where $g$ is a forward scattering constant for the interspin interactions.
More importantly, higher harmonics of Eq.\ \eqref{HaldaneN} generate the terms
of the form~:
\begin{align}
\mathcal{H}_h = %
\sum_{s,s'>0} &G_{ss'}
\int\! dx\, \cos{ \left[ 2(s k_F^\up - s' k_F^\down )x - 2(s\phi_\up - s'\phi_\down )  \right]}  \notag \\
+\sum_{s,s'>0} &\tilde{G}_{ss'} \int\! dx\, \cos{ \left[ 2(s k_F^\up + s'
k_F^\down )x - 2(s\phi_\up + s'\phi_\down )  \right]} \; . \label{H2}
\end{align}
Here $G_{s,s'}$ and $\tilde{G}_{s,s'}$ are (non-universal) amplitudes, and
$k_F^{\sigma} =\pi n_\sigma$ are Fermi momenta. Since the separation of fast
and slow variables is inherent in the bosonization treatment, one has to
discard in \eqref{H2} the terms which oscillate on the lengthscale $\sim
k_F^{-1}$. Strictly speaking, Eq.\ \eqref{H1} is only perturbative in $g$. On
the opposite, Eq.\ \eqref{H0} is assumed to retain its functional form even in
presence of generic same-spin density-density interaction, with both
velocities and Luttinger liquid parameters renormalized by interaction terms
beyond Eq.\ \eqref{H1} and various irrelevant operators, e.g. band curvature
\cite{GiamarchiTheBook}. On a phenomenological level, we can assume Eq.\
\eqref{H0} (where, in general, $K_\sigma\neq 1$)  as coming from an underlying
microscopic model, with Eqs.\ \eqref{H1} and \eqref{H2} regarded as
perturbations.

Equation \eqref{H2} suggests considering generalized commensurabilities of the
form
\begin{equation}
p\,n_\up - q n_\down =0 \;,
\label{pq}
\end{equation}
where $p$ and $q$ are relatively prime integers. Notice that this condition
does not imply the presence of a lattice~: we only require the densities to be
commensurate with each other. Eq.\ \eqref{pq} selects from Eq.\ \eqref{H2} the
terms with $s/s' = p/q$, and the Hamiltonian \eqref{H2} reduces to
%
\begin{equation}
\mathcal{H}_2 = G \int\! dx \, \cos{2\left(p\,\phi_\up (x)-q\phi_\down
(x)\right)} \; ,
\label{H2pq}
\end{equation}
where we only keep the lowest order term, since the scaling dimension of the
operator $\cos{s\phi}$ is $s^2$.
\footnote{For lattice models there is an additional possibility that $p\,n_\up
+ qn_\down = \text{integer}/a_0$, where $a_0$ is the lattice constant. In this
case an additional term of the form $\cos(p\phi_\up + q\phi_\down)$ would
appear in the Hamiltionian. The effects due to such term are similar to those
due to Eq.\ \eqref{H2pq}.}

We now assume that the densities are commensurate via \eqref{pq}, and analyze
the model
$\mathcal{H}=\mathcal{H}_0(\phi_\up)+\mathcal{H}_0(\phi_\down)+\mathcal{H}_1+\mathcal{H}_2$
defined by \eqref{H0}-\eqref{H1}  and \eqref{H2pq}. Since in general this
model is not exactly solvable, the nature of the phases can, in principle, be
determined by an approximate renormalization group (RG) procedure. Rescaling
the fields via $\tilde{\phi}_\up=p\phi_\up$ and
$\tilde{\phi}_\down=q\phi_\down$ the model is brought to the form considered
in Ref.\ \cite{Mathey2007}, where an RG procedure has been carried out including
the renormalization of velocities $v_\sigma$ [see also Ref.\
\cite{PencSolyom1990} in the fermionic language]. For large velocity asymmetry
and strong attractive (repulsive) interactions the system was always found to
collapse (phase separate). Barring such an instability, two regimes were
found, corresponding to the cosine operator \eqref{H2pq} being relevant or
irrelevant in the RG sense.

In the regime where the cosine operator Eq.\ \eqref{H2pq} is irrelevant we are
left with a bilinear Hamiltonian \eqref{H0}--\eqref{H1}, which is diagonalized
by appropriate linear combinations of the fields \cite{Kimura1996}. As a
result one obtains an effective theory $\mathcal{H}_\mathcal{A}$ which
features two decoupled massless fields $\varphi_{1,2}$ with corresponding
velocities $v_{1,2}$ and Luttinger parameters $K_{1,2}$: $
\mathcal{H}_\mathcal{A} = \mathcal{H}_0(\varphi_1) +
\mathcal{H}_0(\varphi_2)$. For an attractive FF mixture, such a theory
describes a 1D analog of the FFLO phase~: all correlations are algebraic in
real space and the pair correlation function oscillates with the FFLO momentum
$Q_\mathrm{FFLO} = | k_F^{\up} - k_F^{\down} |$.

Another regime corresponds to the case
where the cosine in Eq.\ \eqref{H2pq} is relevant in the RG sense. Then the
system has a massive mode $\phi_a$ and a massless mode, $\phi_b$. The effective
theory, $\mathcal{H}_\mathcal{B}$, can be written as~:
\begin{equation}
\mathcal{H}_\mathcal{B} =\mathcal{H}_\mathrm{sG}(\phi_a) +
\mathcal{H}_0(\phi_b)\;,
\label{HB}
\end{equation}
where $\mathcal{H}_\mathrm{sG}(\phi_a)=\mathcal{H}_0(\phi_a) + G\int\!
dx\,\cos2\sqrt{2}\phi_a$ is the sine-Gordon model for the field $\phi_a$. The
Eq.\ \eqref{HB} is characterized by two mode velocities $v_{a,b}$ and two
Luttinger exponents $K_{a,b}$, with $K_a \leqslant 1$, so that $\phi_a$ is
pinned by the minimum of the cosine operator in \eqref{HB}.
Closed-form expressions for the parameters of Eq.\ \eqref{HB} can be easily
obtained in several limiting cases. Indeed, for $v_\up=v_\down$, the exact
eigenmodes of $\mathcal{H} = \mathcal{H}_0(\phi_\up) +
\mathcal{H}_0(\phi_\down) + \mathcal{H}_2$ are~:
\begin{align}
&\phi_a =(p\phi_\up - q\phi_\down)/\sqrt{2}\;, \notag \\
&\phi_b =(qK_\down \phi_\up + pK_\up \phi_\down)/\sqrt{2} \;,
\label{phib}
\end{align}
with the Luttinger exponents 
%
\begin{equation}
K_a = \left( p^2 K_\up +q^2 K_\down \right)/2  \;, \quad\quad%
K_b = K_a K_\up K_\down \; .
%
\label{KaKb}
\end{equation}
%
%
Notice that for higher-order commensurabilities (larger $p$ and $q$) smaller values 
of $K$ are required for $\phi_a$ to acquire a gap, cf. Eq.\ \eqref{KaKb}.

Deep in the massive phase one can make a crude approximation to the
cosine operator in $\mathcal{H}_2$ by replacing it with a mass term $\propto
(p\phi_\up-q\phi_\down)^2$. This leads to:
\begin{align}
& v_b^2 = v_\up v_\down \frac{(p^2 K_\up v_\down + q^2 K_\down v_\up)}{(p^2
K_\up v_\up + q^2 K_\down v_\down)}\; ,
\label{vb} \\
&K_b = \frac{1}{2}K_\up K_\down \sqrt{v_\up v_\down} \frac{(p^2 K_\up + q^2 K_\down)^2}%
{p^2 K_\up v_\down + q^2 K_\down v_\up} \;, \label{Kb}
\end{align}
which reduces to \eqref{KaKb} for $v_\up = v_\down$.

We now turn our attention to an interpretation of the theory
\eqref{HB}, focusing on the novel regime with $p>1$. Obviously, excitations
corresponding to eigenmodes $\phi_{a,b}$ carry both spin and charge. Furthermore, these
excitation correspond to \emph{multiparticle} states in terms of the original
$\up,\down$ particles since a particle of the species $\sigma$ corresponds to
a $2\pi$ kink of the field $\phi_\sigma$ \cite{GiamarchiTheBook}.
To gain further insight to the structure of the massive phase we consider its
correlation properties. We classify
 operators $\mathcal{O}(x)$ according to whether the asymptotic decay
of the correlation functions $\left\langle \mathcal{O}(0)
\mathcal{O}^\dagger(x) \right\rangle$ for $x\to\infty$ is exponential,
$\propto e^{-\lambda_\mathcal{O}|x|}$, or algebraic, $\propto
|x|^{-2\alpha_\mathcal{O}}$. For equal densities, $p=q=1$, the dominant
algebraic order (\textit{i.e.}, the smallest decay exponent
$\alpha_\mathcal{O}$) is found among the two-point operators~: the
superconducting fluctuations, $\mathcal{O}_\mathrm{S}=\Psi_\up \Psi_\down$,
and charge density wave, $\mathcal{O}_\mathrm{CDW} = \sum_{\sigma,\sigma'}
\psi^\dagger_{R\sigma} \delta_{\sigma\sigma'} \psi_{L\sigma'}$, and spin
density wave, $\mathcal{O}^\alpha_\mathrm{SDW} = \sum_{\sigma,\sigma'} \sigma
\psi^\dagger_{R\sigma} {\sigma}_{\sigma\sigma'}^\alpha \psi_{L\sigma'}$. Here
$\psi_{L,R\sigma}$
are left- and right-moving fermions,
respectively, and ${\sigma}^\alpha$ are the
Pauli matrices.

The case $p\neq q$ is markedly different~: First of all, the superconducting
correlations described by $\mathcal{O}_\mathrm{S}$ always decay
\emph{exponentially}, and so do the $x$- and $y$-components of
${\mathcal{O}}^\alpha_\mathrm{SDW}$. For the CDW and SDW$^z$ operators we
write $\mathcal{O}_\mathrm{CDW}=\mathcal{O}_{\mathrm{LR}}^\up +
\mathcal{O}_{\mathrm{LR}}^\down$ and
$\mathcal{O}_\mathrm{SDW}^z=i(\mathcal{O}_{\mathrm{LR}}^\up -
\mathcal{O}_{\mathrm{LR}}^\down )$, where the auxilliary operators
$\mathcal{O}_{\mathrm{LR}}^\sigma=\psi^\dagger_{L\sigma}\psi_{R\sigma}$. Using
\eqref{phib} we find for $x\to\infty$
\begin{equation}
\left\langle \mathcal{O}^\up_\mathrm{LR}(0)\,
\mathcal{O}^{\up}_{\mathrm{LR}}(x)^\dagger \right\rangle  \sim |A(b_\up)|^2 \,
e^{-2ik_F^\up x} \, |x|^{-2\alpha_{\up}} \;,
\label{OLR}
\end{equation}
where $\alpha_\up = q^2 K_b /2 K_a^2$, $b_\up=p K_\up/K_a$ and $A(b_\up)=\left|
\left\langle e^{i\sqrt{2}\phi_a b_\up} \right\rangle \right|^2$. Likewise, for
the $\down$-species the exponent is $\alpha_\down = p^2 K_b/2 K_a^2$ and the
amplitude is $A(b_\down)$ with $b_\down=q K_\down/K_a$. The amplitudes $A(b)$
depend \emph{exponentially} on $b$~: $\log A(b) \propto 1/b^4$
\cite{Lukyanov1997}.
We thus see that correlations of $\mathcal{O}_\mathrm{CDW}$ and
$\mathcal{O}^z_\mathrm{SDW}$ are both given by a superposition of two
power-laws \eqref{OLR} with exponents $\alpha_{\up,\down}$~--- where the
slower the decay, the smaller (\emph{exponentially} smaller) is the
corresponding amplitude.

Given the massive mode in the form \eqref{phib} with $p\neq q$, we construct a
compound operator $\mathcal{O}_{p+q} = \Psi_\down^p \Psi_\up^q$ which has
algebraically decaying correlations. Specializing for the lowest order
commensurability \eqref{pq} with $p=2$ and $q=1$, this corresponds to a
``trimer'' operator $\mathcal{O}_{2+1} = \Psi^\dagger_{\down}
\Psi^\dagger_{\down} \Psi^\dagger_{\up}$. For fermionic $\up$ component the
corresponding decay exponent $\alpha_{2+1} = \left( K_b/2K_a^2 +2 K_a^2/K_b
\right)/2$. We thus see that in this particular case the dominant correlations
in the massive phase are the $2k_F$ density waves \eqref{OLR} for
$\alpha_\up<1/\sqrt{3}$, and the ``trimer'' correlations $\mathcal{O}_{p+q}$
for $\alpha_\up>1/\sqrt{3}$. We stress that the competition between
$\mathcal{O}_\mathrm{LR}$ and $\mathcal{O}_{p+q}$ is generic, in a sense that
it holds irrespective of the statistics of $\up$- and $\down$-particles both
on the lattice and in the continuum.

\textit{Microscopics.---} We now focus on the following question: Is there a
microscopic model whose low-energy effective theory would be given by Eqs.\
\eqref{pq} and \eqref{HB}?

We start from constructing such a model explicitly in the weak-coupling regime
with respect to the interspecies interaction. Namely, for the $\down$
component we take non-interacting fermions (or, equivalently, Tonks bosons) of
the mass $m_\down$ and (linear) density $n_\down$, so that $K_\down=1$ and
$v_\down=\pi n_\down/m_\down$.
For the $\up$ species we take a dipolar Bose gas which is known to be a
Luttinger liquid with $K_\up \to \pi[6\zeta(3)n_\up r_0]^{-1/2}$ as $n_\up
r_0\to\infty$ \cite{Citro2007}. Here $r_0=m_\up d^2/2\pi$ is the effective
Bohr radius associated with the dipole moment $d$ and $\zeta$ is the Riemann
zeta function. We thus see that for $n_\up r_0 = p^4 \pi^2/6\zeta(3)$ we have
$K_\up = 1/p^2$. Furthermore, Galilean invariance fixes the product $v_\up
K_\up = \pi n_\up/m_\up$ \cite{Haldane1981}. Constraining the densities via
\eqref{pq} with $p>q=1$, and assuming $m_\up = p\, m_\down$ we have both
$v_\up=v_\down$ and $p^2 K_\up=q^2 K_\down =1$ \emph{by construction}. Now,
coupling the $\up$ and $\down$ species via, \textit{e.g.,} a short-range
interaction $U \int\! dx\, n_\up(x) n_\down(x)$ with infinitesimal $U$
generates the terms of the form \eqref{H1} and \eqref{H2pq} with $g=U/\pi^2$.
The eigenmodes of the system are then given by Eq.\
\eqref{phib} and a direct calculation yields $K_{a}=p^2 K_\up
\left( 1 + g\frac{K_\up}{v_\up}\frac{p}{4q} \right) + O(g^2)$, and $K_b= p^2
K_\up^2 K_\down \left( 1 - g\frac{K_\up}{v_\up}\frac{p}{4q} \right) + O(g^2)$.
We thus immediately see that having $U<0$ yields $K_a<1$ and hence drives the
system to the gapped phase, where the gap $\Delta$ is exponentially small:
$\ln\Delta \sim -\mathrm{const}/U$. A similar construction can easily be
effected for an FF mixture on a lattice. In this case we take for the $\up$
component, e.g., a model with finite-range interactions
\cite{Gomez-Santos1993}.

In the example above we engineer the theory \eqref{pq} and \eqref{HB} by
coupling a majority of light and non-interacting $\down$ species to the
minority of heavy particles $\up$, which have strong repulsions among
themselves. Such a construction is somewhat \textit{ad hoc}, and requires
fine-tuning. A much more natural alternative is provided by a simple
observation~: even purely local interspecies coupling $U$ generates long-range
effective interactions in higher orders of perturbation theory. Thus, having
finite $U$ \emph{and} $m_\up \neq m_\down$ should be sufficient to divert the
RG flow towards the theory \eqref{HB}. In this case we expect particles of a
light minority component to provide an effective coupling between heavy
particles of majority species.

To this end we consider an asymmetric attractive ($U<0$) Hubbard model
\begin{equation}
H_\mathrm{aH} = -\sum_{\langle ij \rangle\sigma} t_\sigma \left(
c^\dagger_{i,\sigma} c_{j,\sigma} +h.c. \right) +
U\sum_{i}\hat{n}_{i\uparrow}\hat{n}_{i\downarrow} \; ,
\label{aHubb}
\end{equation}
where $c_{i\sigma}$ annihilates a fermion with spin $\sigma$ on a site $i \in
[1,L]$ of a chain lattice of length $L$, $\langle ij \rangle$ stands for pairs
of nearest neighbor sites, $\hat{n}_{i\sigma}=c^{\dagger}_{i\sigma}
c_{i\sigma}$, and $t_\sigma$ are hopping amplitudes for the spin-up and
spin-down components. For $\eta\equiv t_\down/t_\up=1$ the model \eqref{aHubb}
is solvable by Bethe Ansatz techniques even in presence of density imbalance.
For $n_\up \neq n_\down$ the ground state is of the FFLO type \cite{Orso2007},
which, in present language, corresponds to a gapless fixed point theory
$\mathcal{H}_\mathcal{A}$~--- see \cite{ZhaoLiu2008} for a detailed
discussion.

For unequal hopping amplitudes, $\eta \neq 1$, the model \eqref{aHubb} is no
longer integrable, and we resort to numerical simulations using DMRG technique
\cite{DMRG}.
We use lattices of up to $L=80$ sites with open boundary conditions and DMRG
truncation of up to $N_s=400$ states and check that (i) the discarded
probabilities amount to no more than $10^{-7}$, and (ii) the results are
stable with respect to $N_s$. We calculate single-particle density matrices
$\rho_{\sigma}(x) = \langle c_{L/2,\sigma} c_{L/2+x,\sigma}^\dagger \rangle$
and pair-pair correlations $\Gamma(x) = \langle \mathcal{P}_{L/2}
\mathcal{P}_{L/2+x}^\dagger \rangle$, where $\mathcal{P}_{j} = c_{j\up}
c_{j\down}$ is the lattice version of the superconducting operator
$\mathcal{O}_\mathrm{S}$ and $\langle \cdots \rangle$ denotes an expectation
value over the ground state.

\begin{figure}[htb]
\includegraphics[width = 0.95\columnwidth,keepaspectratio=true]{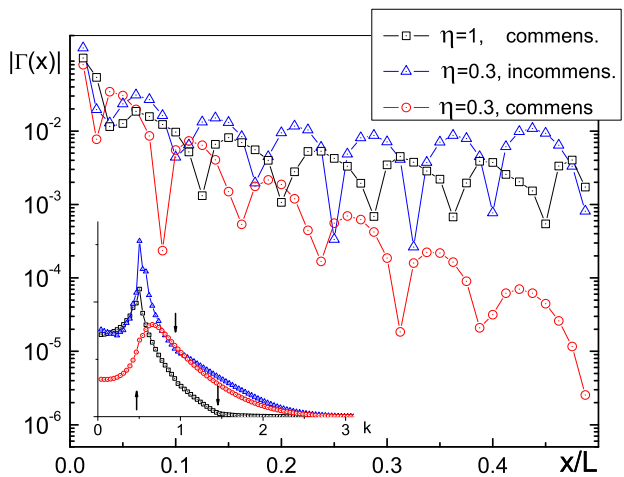}
\caption{Superconducting correlation function $\Gamma(x)$ for the asymmetric
Hubbard model \eqref{aHubb} for $n_\down=2n_\up=3/10$: $\eta=1$ (black
squares) and $\eta=0.3$ (red circles). Shown by blue triangles is $\Gamma(x)$
for $\eta = 0.3$ and `incommensurate' densities $n_\up = 17/80$ and $n_\down =
29/80$. Hubbard coupling is $U=-5t_\up$ and the system size $L=80$. Lines are
guides to the eye. Inset: Fourier transform of $\Gamma(x)$, same color coding.
Arrows indicate the characteristic momenta: $k_F^\down-k_F^\up$, $k_F^\down$
and $k_F^\up + k_F^\down$, respectively, for the `commensurate' densities
$n_\down=2n_\up=3/10$.  We stress that in all these simulations the density
distributions for both components are uniform apart from Friedel oscillations
induced by the open boundary conditions. }
\label{fig:pairpair}
\end{figure}

Fig.\ \ref{fig:pairpair} shows typical results for the pair-pair correlations
$\Gamma(x)$. We find that for small enough hopping asymmetry,
$\eta>\eta_{c1}$, the long-distance decay of both single-particle (not shown)
and two-particle correlations is consistent with the FFLO-type laws
$\Gamma(x)\propto \cos(Q_\mathrm{FFLO}\, x)|x|^{-\gamma}$ and
$\rho_\sigma(x)\propto \cos(k_F^\sigma x) |x|^{-\beta}$. On the contrary, once
the hopping asymmetry exceeds some critical value \emph{and} the densities are
commensurate via \eqref{pq}, the power-law decays change to exponentials,
namely $\Gamma(x)\propto e^{-|x|\lambda} \cos(Q_\mathrm{FFLO}\, x)|x|^{-\gamma^\prime} $ and
likewise for $\rho_\sigma(x)$, thus unequivocally signalling the presence of a
gap. Violating the relation \eqref{pq} destroys the gap, and the correlation
functions decay algebraically again.


The inset in Fig.\ \ref{fig:pairpair}  shows the superconducting  correlation function in momentum space.
Compared to the Hubbard limit ($ \eta=1$), we see that the mass imbalance ($ \eta=0.3$) leads to an overall broadening of the distribution, which now extends well beyond $k_F^\up+k_F^\down$. In addition, the opening of the gap at commensurate filling depletes the superconducting correlation at small momentum.
Detailed investigation of the asymmetric Hubbard model \eqref{aHubb} is beyond
the scope of this Letter and will be reported elsewhere \cite{weOnHubb}.


\textit{Conclusions and outlook.---} Summarizing, we have revealed a generic
mechanism of opening a gap in two-component quantum fluids with
density imbalance in one spatial dimension. The gapped phase appears once
interactions and mass asymmetry between components is strong enough,
\emph{and} the densities satisfy Eq.\ \eqref{pq}. Depending on the microscopic
details, the system develops quasi-long range ordering of either $2k_F$
density waves or of peculiar $(p+q)$-particle composites. The proposed
mechanism applies to mixtures of particles of either statistics, and does not
require the presence of a lattice. Experimental signatures of the proposed
state include (i) the disappearance of the superconducting ordering, and (ii)
appearance of the $(p+q)$-particle composites, which can be detected,
\textit{e.g.}, by noise correlation measurements in the time-of-flight
absorption imaging using the techniques discussed in Ref.\
\cite{KuklovMoritz2007}.

\begin{acknowledgments}
We are indebted to T.~Vekua for illuminating discussions. This work was
supported in part by \textit{Institut Francilien de Recherche sur les Atomes
Froids} (IFRAF) and ANR under grant 08-BLAN-0165-01. G.O. was also supported
by the Marie Curie Fellowship under contract EDUG-038970. Numerical
simulations were performed using DMRG application of the ALPS libraries
\cite{ALPS}.
\end{acknowledgments}


\end{document}